\documentclass[sigconf]{acmart}

\AtBeginDocument{%
  }

\setcopyright{acmlicensed}
\copyrightyear{2026}
\acmYear{2026}
\acmDOI{XXXXXXX.XXXXXXX}

\acmConference[XX '26]{XX 2026}{April XX--
April XX, 2026}{}
\acmISBN{978-1-4503-XXXX-X/26/04}

\usepackage{graphicx}
\usepackage{booktabs}
\usepackage{amsmath}
\usepackage{enumitem}
\usepackage{tcolorbox}
\usepackage{float}

\setlist[itemize]{leftmargin=*, topsep=2pt, itemsep=2pt, parsep=0pt}

\begin{document}

\title{SNAP: A Plan-Driven Framework for Controllable Interactive Narrative Generation}

\author{Geonwoo Bang}
\affiliation{%
  \institution{Sungkyunkwan University}
  \city{Seoul}
  \country{Republic of Korea}}
\email{g7199@g.skku.edu}

\author{DongMyung Kim}
\affiliation{%
  \institution{Sungkyunkwan University}
  \city{Seoul}
  \country{Republic of Korea}}
\email{dmk1304@skku.edu}

\author{Hayoung Oh}
\authornote{Corresponding author}
\affiliation{%
  \institution{Sungkyunkwan University}
  \city{Seoul}
  \country{Republic of Korea}}
\email{hyoh79@skku.edu}

\begin{abstract}
Large Language Models (LLMs) hold great potential for web-based interactive applications, including browser games, online education, and digital storytelling platforms. However, LLM-based conversational agents suffer from spatiotemporal distortions when responding to variant user inputs, failing to maintain consistency with provided scenarios. We propose SNAP (Story and Narrative-based Agent with Planning), a framework that structures narratives into Cells with explicit Plans to prevent narrative drift in web environments. By confining context within each Cell and employing detailed plans that specify spatiotemporal settings, character actions, and plot developments, SNAP enables coherent and scenario-consistent dialogues while adapting to diverse user responses. Via automated and human evaluations, we validate SNAP's superiority in narrative controllability, demonstrating effective scenario consistency despite variant user inputs in web-based interactive storytelling.
\end{abstract}


\maketitle

\begin{figure}[!htbp]
\centering
  \includegraphics[width=\columnwidth]{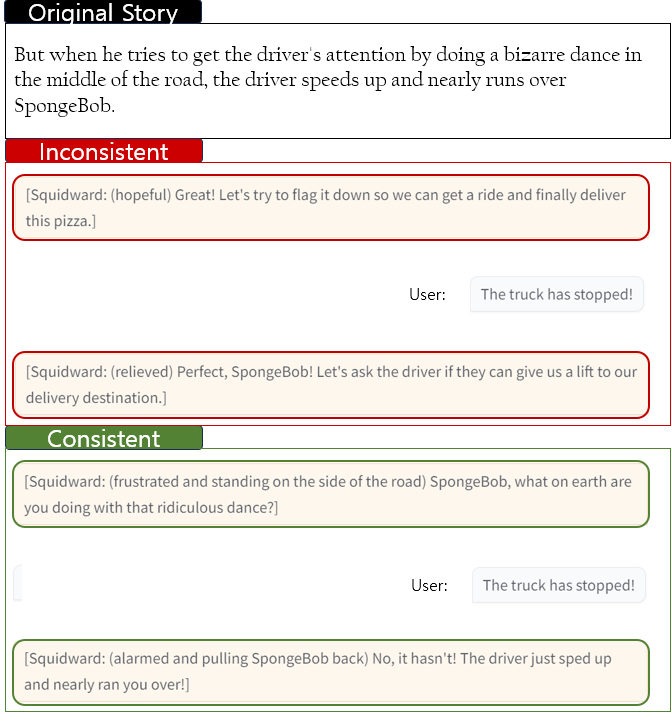}
  \caption{(Top) Squidward deviates from the original storyline due to the user's input. (Bottom) Squidward remains aligned with the original storyline despite the user's attempt to alter it.}
  \label{fig:fig1}
\end{figure}

\begin{figure*}[!htbp]
\centering
  \includegraphics[width=0.9\textwidth]{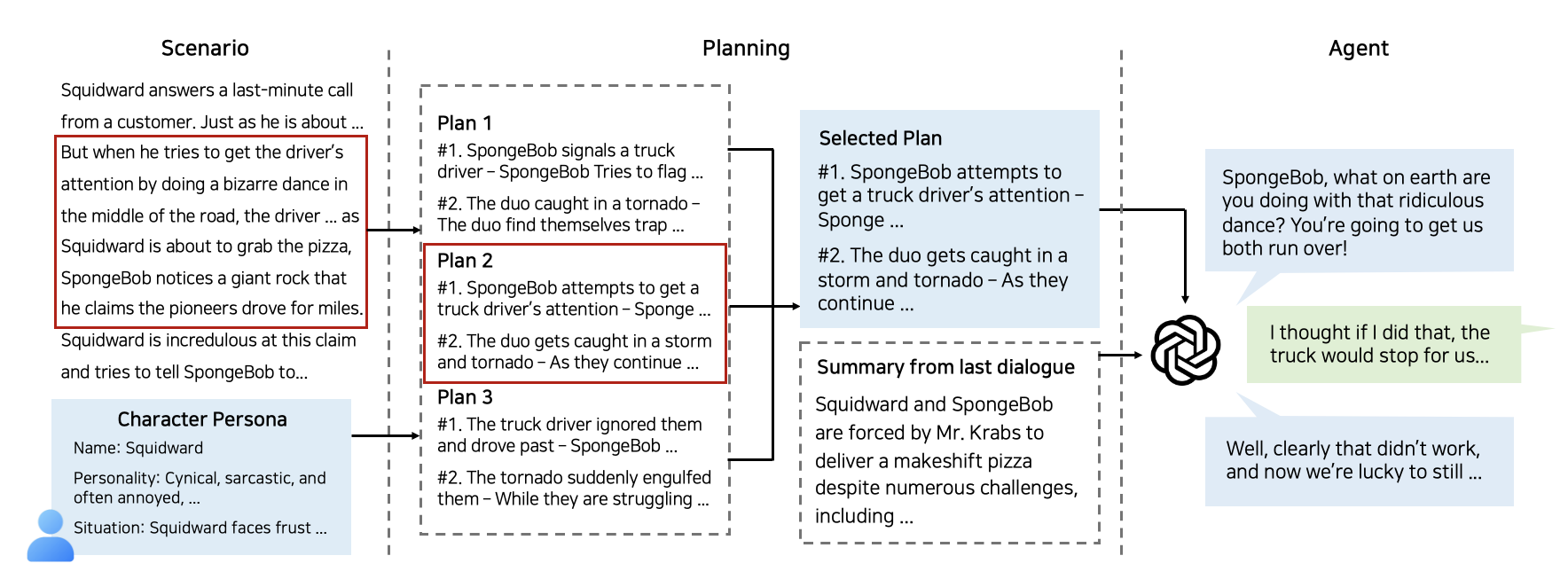}
  \caption{Cell-based narrative structure showing inputs (story segment, character personas, previous summary, user inputs) and processing flow.}
  \label{fig:framework}
\end{figure*}

\section{Introduction}

Recent advances in LLM agents have demonstrated impressive capabilities in simulation \cite{park2023generativeagentsinteractivesimulacra,Argyle_2023} and conversation. However, these systems face challenges in maintaining consistency over extended interactions. Character-specific approaches have emerged: protective scenes \cite{shao-etal-2023-character} prevent out-of-character responses, while TimeChara \cite{ahn-etal-2024-timechara} addresses temporal consistency. Memory-augmented agents further stabilize persona over time \cite{10.1145/3613905.3650839}.

Yet a critical problem persists: spatiotemporal distortion. When users ask questions beyond the current scene boundary (probing future events or distant locations), LLMs with full narrative context readily answer, breaking temporal causality and spatial coherence. A character might reference events that haven't occurred yet, or describe places they shouldn't know about, destroying narrative immersion. Unexpected user turns can derail pacing, corrupt character behavior, and erode creator intent. Rule-based guardrails break fluency, prompt tweaks fade over multi-turn dialogue, and costly fine-tuning struggles to generalize across narratives. See Figure~\ref{fig:fig1}

Controllable generation research has explored fine-grained controllable story models \cite{wang-etal-2022-chae}, grounded dialogue generation to curb hallucination \cite{yang-etal-2023-refgpt}, and planning-based tool-use methods \cite{yao2023reactsynergizingreasoningacting}. But most work focuses on single-turn generation or document-level control and does not enforce scene-level narrative steering in interactive dialogue.

SNAP (Story and Narrative-based Agent with Planning) centers controllability on two core ideas. First, Cell-based segmentation breaks down narrative scope into manageable scenes, reducing the burden on LLM context while keeping each segment adjustable. Critically, by limiting context to the current Cell, the agent naturally lacks information about future events, preventing spatiotemporal leakage. Second, explicit plans per Cell spell out required events, actions, and constraints while leaving surface-level conversation free. Plans guide rather than dictate, keeping dialogues natural yet oriented to the plot.

Unlike existing approaches that focus on character consistency, SNAP addresses narrative control, ensuring plot progression aligns with creator intent even when characters remain consistent. SNAP extends planning-based control to multi-turn interactive dialogues where user inputs continuously perturb the generation process, requiring dynamic redirection rather than static control.

\section{SNAP Framework}

\begin{figure}[t]
\centering
  \includegraphics[width=1.02\columnwidth]{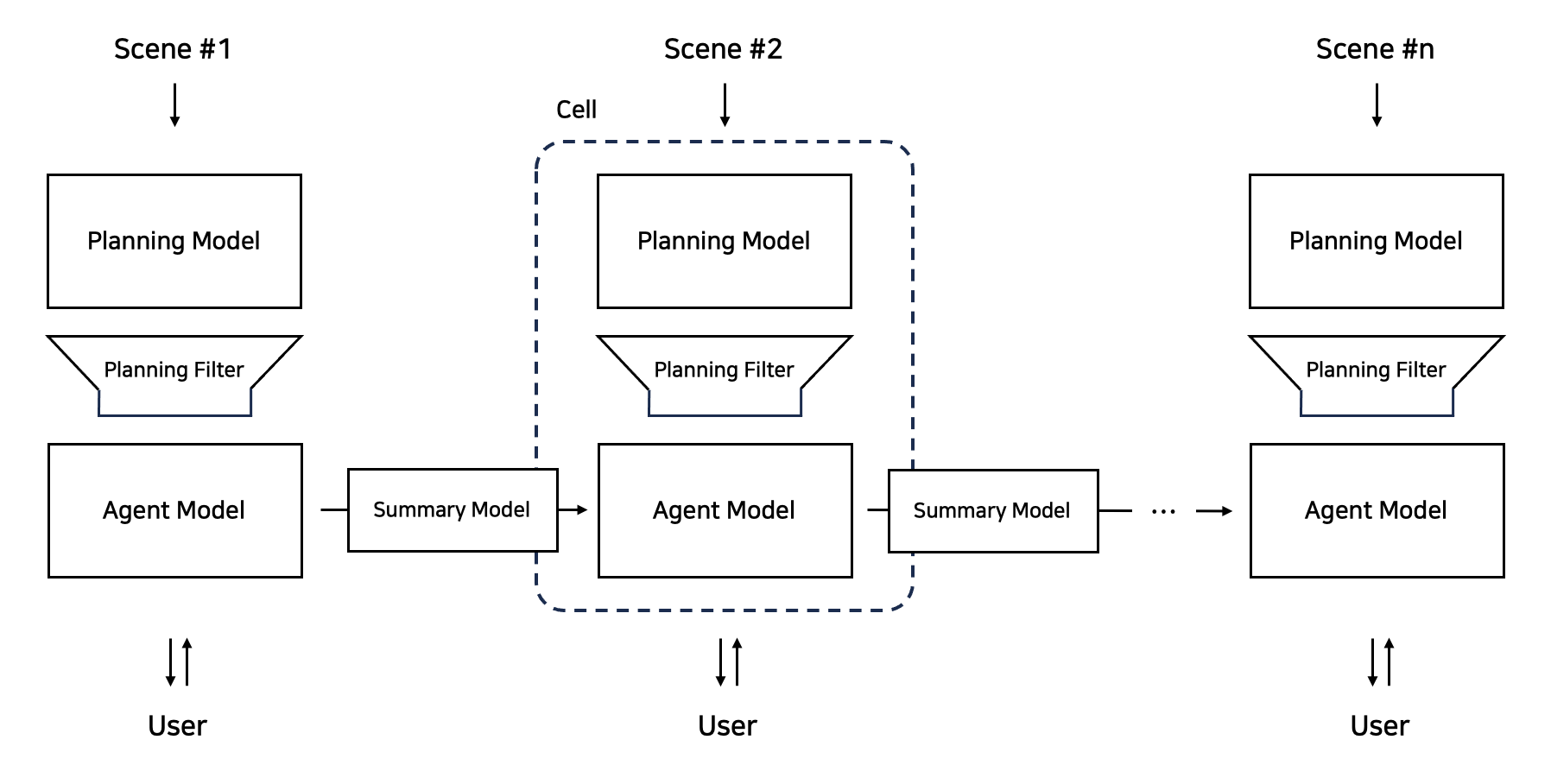}
  \caption{SNAP Framework Architecture. Narratives are segmented into Cells, each processed through Planning, Filtering, Dialogue Agent, and Summary.}
  \label{fig:framework}
\end{figure}

\subsection{Cell-Based Narrative Segmentation}

A \textit{Cell} represents a self-contained narrative unit with clear boundaries. Each Cell receives: \textbf{Story segment} (current scene, 10 sentences in this work), \textbf{Character personas} (traits, roles, background), \textbf{Previous summary} (condensed prior context), and \textbf{User inputs} (real-time conversational turns). Figure~\ref{fig:framework} illustrates this architecture.

\noindent \textbf{Why Cells Enables Efficient Control}: By establishing clear scene boundaries, Cells provide:
\begin{enumerate}[leftmargin=2em]
\item \textit{Scope limitation}: LLMs split long context into smaller cells for stepwise management
\item \textit{Progress tracking}: Completion of a Cell marks concrete narrative advancement
\item \textit{Intervention points}: Creators can adjust Plans between Cells based on user behavior
\end{enumerate}

\subsection{Plan-Driven Dialogue Generation}

\subsubsection{Planning Model}

For each Cell, a Planning Model (GPT-4o \cite{achiam2023gpt4} with $T=0.3$) generates candidate Plans. A Plan consists of sequential subplans. Each Plan lists sequential subplans that spell out \emph{Objective}-what should happen in this substep, \emph{Details}-locations, items, character actions required, and \emph{Constraints}-information that should/shouldn't be revealed yet.

\subsubsection{Planning Filter}

To ensure Plan quality, we generate multiple candidates ($n=5$) and select the optimal one using a scoring function combining three criteria. Weights are derived from PCA on random selected examples ($k=500$).
\begin{equation}
S(P_i) = 0.289 \cdot \text{Coh}(P_i) + 0.354 \cdot \text{Con}(P_i) + 0.357 \cdot \text{Per}(P_i)
\end{equation}

\begin{itemize}[label={}]
\item \textbf{Coherence}: Semantic alignment with story segment (BERTScore~\cite{zhang2020bertscoreevaluatingtextgeneration})
\item \textbf{Connectivity}: Logical flow between subplans (G-eval~\cite{liu-etal-2023-g})
\item \textbf{Personality}: Character consistency (G-eval)
\end{itemize}

\medskip

The optimal plan \(P_{\text{best}}\) is selected as the one with the maximum score through equation~\ref{eq:maximum}.
\begin{equation}
  P_{\text{best}} = \arg\max_{P_i} S(P_i)
  \label{eq:maximum}
\end{equation}

\subsubsection{Agent Model}

Given both the optimal plan $P_{\text{best}}$ and the summary of the previous cell's dialogue, the Agent Model (GPT-4o with $T = 0.3$) generates dialogue by balancing two objectives:

\begin{enumerate}[leftmargin=1.5em]
\item \emph{Conversational naturalness}: Respond appropriately to user
\item \emph{Narrative progression}: Advance toward Plan objectives
\end{enumerate}

When users deviate from the intended narrative, the agent employs a three-step engagement strategy:
\begin{enumerate}[leftmargin=1.5em]
    \item \emph{Acknowledge} the user input to maintain engagement
    \item \emph{Respond briefly} with contextually relevant information
    \item \emph{Redirect strategically} back to the current objective
\end{enumerate}

This approach ensures players feel heard while maintaining narrative coherence. Crucially, Cell-based segmentation enables effective redirection: by confining context to short scene segments, the agent lacks information about distant plot points and naturally recognizes off-topic queries as outside its current scope. In contrast, long-context scenarios allow agents to access future events prematurely, causing spatiotemporal distortions when users probe beyond the intended narrative boundary.
\vspace{0.3cm}

\noindent\textbf{Example Interaction:}

\begin{tcolorbox}[colback=gray!10, colframe=gray!50, arc=2mm, boxrule=0.5pt]
\noindent\textbf{Agent:} ``SpongeBob, we need to get on the boat and start the delivery right away!''

\vspace{0.1cm}

\noindent\textbf{Player:} ``Why don't we just take a taxi?'' \textit{[off-topic query]}

\vspace{0.2cm}

\noindent\textbf{Agent:} ``Mr. Krabs wouldn't want to spend extra money on a taxi, so we're stuck with the boat.''

\vspace{0.1cm}
\noindent\textit{[acknowledges concern + uses character motivation + redirects to delivery task]}
\end{tcolorbox}

\vspace{0.1cm}

\noindent The agent progressively follows the narrative plan, advancing through subplots sequentially. Upon completing a story segment, the system emits an \texttt{EOD} (End of Dialogue) marker to signal transition to the next narrative cell.

\subsection{Summary Model and Cell Transitions}

At cell completion, a Summary Model condenses the dialogue into key events, character emotional states, and unresolved elements, providing continuity for the next cell while preventing context overflow. We used BART trained on DialogSum~\citep{lewis-etal-2020-bart, chen-etal-2021-dialogsum}.

\section{Experiments}

We evaluate SNAP on Wikiplot, a dataset containing over 112K narrative plots. We randomly sample 13 plots and segment each into cells of approximately 10 sentences. Prompts for each phase are automatically generated using GPT-4o by providing a description of the target object.

\subsection{Experimental Setup}

\noindent\textbf{Baselines:} (1) \emph{Vanilla GPT-4o}: Standard conversation using only prompt engineering, (2) \emph{SNAP w/o Cell}: Agent with explicit planning but without cell segmentation.

\medskip

\noindent\textbf{Evaluation Metrics:}
We assess narrative controllability across four key dimensions on 1-5 scale: \emph{Continuity} (smooth temporal and spatial narrative flow), \emph{Information Appropriateness} (correct timing of information revelation), \emph{Non-Redundancy} (avoidance of repetitive dialogue), and \emph{Linearity} (ability to maintain narrative direction after user diversions).

\subsection{Results}

\noindent\textbf{Automatic Evaluation.} We conduct automatic evaluation using two character agents conversing within narrative scenarios, with a separate GPT-4o model as G-eval evaluator. Scenario names are specified in column headers, and Continuity / Information Appropriateness / Non-Redundancy / Linearity scores are recorded per cell. Table~\ref{table:scenario} presents the comparison across 13 scenarios.

\begin{table}[!htbp]
\centering
\caption{Scenario-level automatic evaluation (13 scenarios, higher is better)}
\label{table:scenario}
\footnotesize
{\setlength{\tabcolsep}{3pt}%
\begin{tabular}{lcccccccc}
\toprule
\textbf{Scenario} & \multicolumn{2}{c}{\textbf{Cont.}} & \multicolumn{2}{c}{\textbf{Non-Red.}} & \multicolumn{2}{c}{\textbf{Info.}} & \multicolumn{2}{c}{\textbf{Lin.}} \\
\cmidrule(lr){2-3} \cmidrule(lr){4-5} \cmidrule(lr){6-7} \cmidrule(lr){8-9}
 & SNAP & Vanilla & SNAP & Vanilla & SNAP & Vanilla & SNAP & Vanilla \\
\midrule
Buffy & \textbf{4.738} & 3.663 & \textbf{3.738} & 3.275 & \textbf{3.588} & 2.475 & 4.925 & \textbf{4.988} \\
Simpsons & \textbf{5.000} & 4.325 & \textbf{3.525} & 2.050 & \textbf{3.650} & 2.325 & 4.950 & \textbf{4.975} \\
Blackmail & \textbf{4.883} & 3.550 & \textbf{3.650} & 2.700 & \textbf{3.517} & 2.300 & 4.950 & \textbf{4.983} \\
Atlantis & \textbf{4.610} & 3.600 & \textbf{3.860} & 3.220 & \textbf{3.390} & 2.380 & 4.870 & \textbf{4.990} \\
Jargon & \textbf{4.508} & 3.742 & \textbf{3.875} & 3.183 & \textbf{3.325} & 2.450 & 4.725 & \textbf{4.842} \\
Tootle & \textbf{4.571} & 3.793 & \textbf{3.750} & 3.293 & \textbf{3.207} & 2.514 & 4.764 & \textbf{4.864} \\
Runaway & \textbf{4.625} & 3.938 & \textbf{3.906} & 3.381 & \textbf{3.425} & 2.600 & 4.794 & \textbf{4.881} \\
Pinky & \textbf{4.228} & 3.967 & \textbf{3.794} & 3.450 & \textbf{3.261} & 2.628 & 4.706 & \textbf{4.894} \\
Kabir & \textbf{4.260} & 3.970 & \textbf{3.815} & 3.505 & \textbf{3.265} & 2.615 & 4.715 & \textbf{4.905} \\
Klissan & \textbf{4.245} & 3.732 & \textbf{3.800} & 3.550 & \textbf{3.214} & 2.559 & 4.741 & \textbf{4.914} \\
Pizza & \textbf{4.296} & 3.763 & \textbf{3.763} & 3.504 & \textbf{3.333} & 2.533 & 4.679 & \textbf{4.892} \\
Moon & \textbf{4.346} & 3.796 & \textbf{3.858} & 3.535 & \textbf{3.385} & 2.519 & 4.704 & \textbf{4.900} \\
Scarlett & \textbf{4.321} & 3.807 & \textbf{3.868} & 3.575 & \textbf{3.357} & 2.550 & 4.654 & \textbf{4.907} \\
\midrule
\textbf{Avg} & \textbf{4.510} & 3.819 & \textbf{3.800} & 3.171 & \textbf{3.378} & 2.496 & 4.783 & \textbf{4.910} \\
\bottomrule
\end{tabular}%
}
\end{table}

\medskip

\noindent\textbf{Ablation Study.} Table~\ref{table:cell} shows the impact of Cell segmentation on narrative quality.

\begin{table}[!htbp]
\centering
\caption{Cell mechanism ablation: average across 13 scenarios}
\label{table:cell}
\small
\scalebox{0.9}{
\begin{tabular}{lcccc}
\toprule
\textbf{Method} & \textbf{Cont.} & \textbf{Non-Red.} & \textbf{Info.} & \textbf{Lin.} \\
\midrule
SNAP & \textbf{4.510} & \textbf{3.800} & \textbf{3.378} & 4.783 \\
SNAP w/o Cell & 3.714 & 3.040 & 3.546 & 4.850 \\
Vanilla & 3.819 & 3.171 & 2.496 & \textbf{4.910} \\
\bottomrule
\end{tabular}
}
\end{table}

\noindent\textbf{Human Evaluation.} We conducted evaluation with 23 university students in a counterbalanced design. Participants played SpongeBob in a pizza delivery scenario, conversing with Squidward (the agent). See Table~\ref{table:human}.

\begin{table}[!htbp]
\centering
\caption{Human evaluation results (n=23, paired). All differences significant at $p<0.01$.}
\label{table:human}
\small
\scalebox{0.9}{
\begin{tabular}{lccc}
\toprule
\textbf{Metric} & \textbf{SNAP} & \textbf{Vanilla} & \textbf{$\Delta$\%} \\
\midrule
Continuity & \textbf{3.913} & 2.478 & +57.7\% \\
Info Appropriateness & \textbf{4.043} & 2.826 & +42.8\% \\
Non-Redundancy & \textbf{4.087} & 2.478 & +65.0\% \\
Linearity & \textbf{4.391} & 3.261 & +34.6\% \\
\bottomrule
\end{tabular}
}
\end{table}

\subsection{Research Questions}

\noindent\textbf{RQ1: Does SNAP effectively mitigate spatiotemporal distortions in interactive narratives?}

Table~\ref{table:scenario} demonstrates SNAP's effectiveness across automated scenarios. SNAP consistently outperforms Vanilla GPT-4o in metrics directly measuring spatiotemporal consistency: Continuity (4.510 vs 3.819, +18.1\%) and Information Appropriateness (3.378 vs 2.496, +35.3\%). These improvements indicate that SNAP's structural mechanisms (Cell-based segmentation and Plan-driven generation) effectively prevent the hallucinations and timeline inconsistencies that emerge when LLMs generate responses to variant user inputs without narrative constraints.

\medskip

\noindent\textbf{RQ2: Do users perceive improved narrative coherence in real interactions?}

Table~\ref{table:human} presents human evaluation results from 23 participants engaging in interactive storytelling. SNAP demonstrates statistically significant improvements across all controllability dimensions ($p<0.01$): Continuity (+57.7\%), Information Appropriateness (+42.8\%), Non-Redundancy (+65.0\%), and Linearity (+34.6\%). Qualitative feedback corroborates these quantitative findings: participants reported that Vanilla agents exhibited \textit{"timeline confusion"} and \textit{"started too late in the story"}, while SNAP maintained \textit{"appropriate pacing"} and \textit{"successful redirection"}. This convergence between objective metrics and subjective user experience validates that SNAP's structural mechanisms translate into perceptible improvements in narrative quality.

Open-ended qualitative feedback from participants are:

\begin{tcolorbox}[colback=gray!10, colframe=gray!50, arc=2mm, boxrule=0.5pt, title=\textbf{Qualitative Feedback from Participants}]

\textbf{SNAP Strengths:} \textit{"Progressed at more appropriate pace, didn't stall"} · \textit{"Successfully redirected back when I went off-topic"} · \textit{"Fewer repetitive responses, story kept moving"}

\vspace{0.1cm}

\textbf{Vanilla Failures:} \textit{"Started too late in the story timeline"} · \textit{"Got stuck repeating same statements"} · \textit{"Followed my diversions, lost the original plot"}

\vspace{0.1cm}

\textbf{SNAP Limitations:} \textit{"Sometimes felt restrictive, less freedom to explore"}

\end{tcolorbox}

\medskip

\noindent\textbf{RQ3: What is the role of Cell-based segmentation in controllable generation?}

The ablation analysis in Table~\ref{table:cell} reveals that Cell-based segmentation serves as a critical structural component. While SNAP w/o Cell retains Plan-driven generation, removing Cells causes performance to approach Vanilla levels in Continuity (3.714 vs Vanilla 3.819) and fall below SNAP in Non-Redundancy (3.040 vs SNAP 3.800). Notably, automatic evaluation shows marginally higher Linearity for systems without Cells (Vanilla: 4.910, SNAP w/o Cell: 4.850 vs SNAP: 4.783), yet human evaluation dramatically reverses this pattern (SNAP: 4.39 vs Vanilla: 3.26, +34.6\%). This discrepancy indicates that Cell boundaries provide essential structural guidance for handling unpredictable user diversions—a capability that emerges only under authentic interactive conditions rather than cooperative agent simulations. The trade-off manifests as occasional user-perceived restrictions, reflecting the inherent tension between structural control and exploratory freedom.

\section{Key Findings}

Our experiments demonstrate three key findings. First, Plans are the primary control mechanism: the Plan-based approach shows substantial improvements over Vanilla even without Cell segmentation, with the largest gains in Information Appropriateness (+31.8\%), validating that explicit Plans enable proper narrative pacing. Second, Cells enhance scalability by preventing the degradation seen in Vanilla agents as context grows, proving crucial for extended narratives. Third, human evaluation reveals true controllability: while automatic evaluation showed marginally lower Linearity for SNAP, real user interactions demonstrate a 34.6\% improvement (4.39 vs 3.26). This validates that SNAP's control mechanisms, which appear as overhead in automatic settings, effectively handle real user diversions, directly fulfilling its core purpose of maintaining narrative direction.

\section{Conclusion}

In this paper, we identified the controllability problem in LLM-based interactive narrative systems where agents must maintain natural conversational engagement while preserving creator-specified narrative coherence despite unpredictable user inputs. We proposed SNAP, a novel framework that employs Cell-based segmentation to provide structural boundaries and Plan-driven generation to encode creator intent, successfully addressing the fundamental tension between conversational flexibility and narrative control. As LLMs become increasingly integrated into web-based interactive systems, frameworks like SNAP that balance user adaptation with narrative control will be essential for production deployment in browser games, educational platforms, and interactive fiction applications. Our work provides a foundation for controllable narrative generation, opening paths toward more reliable, creator-aligned interactive experiences on the web.

\bibliographystyle{ACM-Reference-Format}
\bibliography{base}

\end{document}